\documentclass[aps,showpacs,epsfig,twocolumn]{revtex4}
\usepackage{epsfig}
\usepackage{amsmath}

\begin{document}

\title{\textbf{{\Large {Normal metal to ferromagnetic superconductor tunneling}}}}
\author{Naoum Karchev and Tzanko Ivanov\cite{byline}}

\begin{abstract}
We study the point-contact tunneling between a normal metal and a
ferromagnetic superconductor. In the case of magnon-induced pairing the
tunneling conductance is a continuous and smooth function of the applied
voltage. For small values of the applied voltage the Ohm law holds. We show
that one can obtain the magnetization and the superconducting order
parameter from the tunneling conductance. In the case of paramagnon-induced
superconductivity the tunneling does not depend on the magnetization. We
argue that the tunneling experiment can unambiguously determine the correct
pairing mechanism in the ferromagnetic superconductors.
\end{abstract}

\maketitle

\affiliation{Department of Physics, University of Sofia, 1126 Sofia,
Bulgaria}

\affiliation{Department of Physics, University of Sofia, 1126 Sofia,
Bulgaria}

The discovery of unconventional superconductivity caused an explosive growth
of activities in various fields of condensed-matter research, stimulating
studies of the basic mechanisms leading to this phenomenon. The most direct
way to identify the Cooper pairs is from measurements of their spin
susceptibility, which can be determined by the Knight shift, from
measurements of nuclear spin-lattice relaxation rate $1/T_1$, probed by
nuclear magnetic resonance and nuclear quadrupole resonance, and by electron
tunneling.

In conventional superconductors, the quasi-particles form Cooper pairs in a
spin-singlet state which has zero total spin. The existence of the gap in
the quasi-particle spectrum leads to unusual properties of the systems: i)
The specific heat decreases exponentially at low temperature, as opposed to
the linear temperature dependence in the Fermi liquid theory. ii)The normal
metal superconductor tunneling experiments show that the electrons from the
normal side of the junction can tunnel through and become an excited
quasi-particle on the superconducting side if the applied voltage is larger
then the gap\cite{tunn}. All these properties are well understood on the
basis of the BCS theory of superconductivity.

The discovery of superconductivity in a single crystal of $UGe_2$\cite{tunn1}%
, $URhGe$\cite{tunn2} and $ZrZn_2$\cite{tunn3} revived the interest in the
coexistence of superconductivity and ferromagnetism. The experiments
indicate that the superconductivity is confined to the ferromagnetic phase,
the ferromagnetic order is stable within the superconducting phase (neutron
scattering experiments), and the specific heat anomaly associated with the
superconductivity in these materials appears to be absent. The specific heat
depends on the temperature linearly at low temperature.

At ambient pressure $UGe_2$ is an itinerant ferromagnet below the Curie
temperature $T_c=52K$, with low-temperature ordered moment of $%
\mu_s=1.4\mu_B/U$. With increasing pressure the system passes through two
successive quantum phase transition, from ferromagnetism to ferromagnetic
superconductivity at $P\sim 10$ kbar, and at higher pressure $P_c\sim$ 16
kbar to paramagnetism\cite{tunn1,tunn4}. At the pressure where the
superconducting transition temperature is a maximum $T_{sc}=0.8K$, the
ferromagnetic state is still stable with $T_c=32K$. The survival of bulk
ferromagnetism below $T_{sc}$ has been confirmed directly via elastic
neutron scattering measurements\cite{tunn4}. The specific heat coefficient $%
\gamma=C/T$ increases steeply near 11 kbar and retains a large and nearly
constant value\cite{tunn5}.

Specifically, $UGe_2$ has strong spin orbit-interaction that leads to an
unusually large magneto-crystalline anisotropy with an easy magnetization
axis along the shortest crystallographic axis.

The ferromagnets $ZrZn_2$ and $URhGe$ are superconducting at ambient
pressure with superconducting critical temperatures $T_{sc}=0.29K$\cite
{tunn3} and $T_{sc}=0.25K$\cite{tunn4} respectively. $ZrZn_2$ is
ferromagnetic below the Curie temperature $T_c=28.5K$ with low-temperature
ordered moment of $\mu_s=0.17\mu_B$ per formula unit, while for $URhGe$\,\,$%
T_c=9.5K$ and $\mu_s=0.42\mu_B$. The low Curie temperatures and small
ordered moments indicate that compounds are close to a ferromagnetic quantum
critical point.

We shall discuss two mechanisms of Cooper pairing in ferromagnetic metals:
superconductivity induced by longitudinal spin fluctuations\cite{tunn6} and
magnon exchange mechanism of superconductivity\cite{tunn7}. In the case of
paramagnon induced superconductivity\cite{tunn6} the order parameters are
spin parallel components of the spin triplet. The theory predicts that spin
up and spin down fermions form Cooper pairs, and hence the specific heat
decreases exponentially at low temperature. The phenomenological theories 
\cite{tunn8} circumvent the problem assuming that only majority spin
fermions form pairs, and hence the minority spin fermions contribute the
asymptotic of the specific heat. The magnon exchange mechanism of
superconductivity was developed\cite{tunn7} to explain in a natural way the
fact that the superconductivity in $UGe_2$, $ZrZn_2$ and $URhGe$ is confined
to the ferromagnetic phase.The order parameter is a spin anti-parallel
component of a spin-1 triplet with zero spin projection ($%
\uparrow\downarrow+\downarrow\uparrow$). The onset of superconductivity
leads to the appearance of Fermi surfaces in the spin up and spin down
momentum distribution functions. As a result, the specific heat depends on
the temperature linearly, at low temperature.

During the last years tunneling spectroscopy has been applied to identify
the pairing symmetry. The most basic idea of tunneling spectroscopy was
first proposed by Bardeen\cite{tunn9} who introduced the tunnel Hamiltonian
approximation for describing a tunnel junction. The concept of the tunneling
Hamiltonian\cite{tunn10} became universally adopted for the discussion of
tunneling in superconductors. The idea is to write the Hamiltonian as a sum
of three terms 
\begin{equation}
H\,=\,H_{L}+H_{R}+H_{T}  \label{tunn1}
\end{equation}
The first two terms $H_{L}$ and $H_{R}$ are considered to be independent,
expressed in terms of two independent sets of Fermi operators $%
c_{k,\sigma},c^+_{k,\sigma}$ and $d_{q,\sigma}, d^+_{q,\sigma}$, where $%
\sigma=(\uparrow,\downarrow)$. Tunneling is caused by the term $H_{T}$ 
\begin{equation}
H_{T}\,=\,\sum\limits_{k,q,\sigma}\left[t_{k,q}c^+_{k,\sigma}d_{q,\sigma}\,+%
\, t^{\ast}_{k,q}d^+_{q,\sigma}c_{k,\sigma}\right]  \label{tunn2}
\end{equation}
The tunneling matrix element $t_{k,q}$ describes the transfer of the
particle through the junction. In this letter we focus our attention on the
point contact case, therefore it is assumed $t_{k,q}$ to depend only on the
wave vectors on the two sides $"k"$ and $"q"$. The tunneling takes place
over a very narrow span of energies near the Fermi surfaces of the "left"
and "right" systems, that is why it is an adequate approximation to treat
the transfer rate $t_{k,q}$ as a constant which is evaluated at $k_F$ and $%
q_F$ $(t_{q_F,k_F}=t^\ast_{q_F,k_F}=t)$.

We consider the tunneling process from a normal metal to a ferromagnetic
superconductor. $H_{L}$ is the Hamiltonian of the system of free spin 1/2
fermions with dispersion $\epsilon _{k}=\frac{k^{2}}{2m}-\mu $\thinspace , $%
H_{L}\,=\,\sum\limits_{k,\sigma }\epsilon _{k}c_{k\sigma }^{+}c_{k\sigma }$,
and $H_{R}$ is the Hamiltonian for the ferromagnetic superconductor. For the
magnon-induced superconductivity it has the form 
\begin{eqnarray}
& & H_{R}=\sum\limits_{k,\sigma }\epsilon _{k\sigma }d_{k\sigma
}^{+}d_{k\sigma }+  \label{tunn3} \\
& & \sum\limits_{k}\Delta _{k}\left[ d_{k\uparrow }d_{-k\downarrow
}+d_{k\downarrow }d_{-k\uparrow }+d_{-k\uparrow }^{+}d_{k\downarrow
}^{+}+d_{-k\downarrow }^{+}d_{k\uparrow }^{+}\right]  \notag
\end{eqnarray}
while for paramagnon-induced superconductivity it is 
\begin{equation}
H_{R}=\sum\limits_{k,\sigma }\epsilon _{k\sigma }d_{k\sigma }^{+}d_{k\sigma
}+\frac{1}{2}\sum\limits_{k}\Delta _{k}\left[ d_{k\uparrow }d_{-k\uparrow
}+d_{-k\uparrow }^{+}d_{k\uparrow }^{+}\right] .  \label{tunn4}
\end{equation}
In Eqs.(\ref{tunn3},\ref{tunn4}) $\epsilon _{k\uparrow }=\frac{k^{2}}{2m}%
-\mu -H$ and $\epsilon _{k\downarrow }=\frac{k^{2}}{2m}-\mu +H$, where $H$
is the energy of the Zeeman splitting, which is proportional to the low
temperature ordered moment. There is a symmetry relation for the gap
function $\Delta _{k}$ which follows from the anti-commutation of spin $1/2$
fermions $\Delta _{-k}=-\Delta _{k}$. The gap has the form $\Delta
_{k}=\Delta _{0}\cos \theta _{k}$\cite{tunn6,tunn7}.

The strong spin-orbit coupling in $UGe_2$ and $URhGe$ requires pseudo-spin
technique\cite{tunn10a} . The spin rotations, in that case, are accompanied
by a rotation in momentum space. Hence if we consider tunneling process from
a normal metal to a ferromagnetic superconductor, the tunneling Hamiltonian
is invariant if and only if the tunneling matrix element is a constant. This
approximation is widely accepted, but in our case it is crucial for the
applicability of the tunneling Hamiltonian.

Next we calculate the current through the tunnel junction. It is defined as 
\begin{equation}
I=-e\left\langle \frac{d}{d\tau }\sum_{k,\sigma }c_{k,\sigma
}^{+}c_{k,\sigma }\right\rangle .
\end{equation}
The tunneling current can be presented in the form 
\begin{equation}
I=et\sum_{k,q,\sigma }\left[ \ll d_{q\sigma },c_{k\sigma }^{+}\gg _{<}\left(
\tau ,\tau \right) -\ll c_{k\sigma },d_{q\sigma }^{+}\gg _{<}\left( \tau
,\tau \right) \right]  \label{current}
\end{equation}
where we have introduced the Green function in the Keldysh representation $%
\ll A,B^+\gg _{<}\left( \tau _{1},\tau _{2}\right) =i\left\langle B^{+}\left(
\tau _{2}\right) A\left( \tau _{1}\right) \right\rangle $\cite{tunn11}. Let
us consider first the magnon-induced superconductivity. We want to compute
the tunneling current to the lowest order $\left( \sim t^{2}\right) $ in $t$%
\cite{tunn12}. To that goal, we use the equations of motions of the Green
functions in Eq. (\ref{current}) to cast the current into the form 
\begin{eqnarray}
I &=&et^{2}\sum\limits_{k,q,\sigma }\int \frac{d\omega }{2\pi }[\left[
\Sigma _{\sigma r}\left( q,\omega \right) -\Sigma _{\sigma a}\left( q,\omega
\right) \right] G_{\sigma <}\left( k,\omega \right) -  \notag \\
&&\Sigma _{\sigma <}\left( q,\omega \right) \left[ G_{\sigma r}\left(
k,\omega \right) -G_{\sigma a}\left( k,\omega \right) \right] ]
\end{eqnarray}
where 
\begin{eqnarray}
\Sigma _{\uparrow \mu }(q,\omega ) & = & u_{q}^{2}A_{1\mu }(q,\omega)
+v_{q}^{2}A_{2\mu }^{+}(q,\omega ), \\
\Sigma _{\downarrow \mu }(q,\omega ) & = & v_{q}^{2}A_{1\mu }^{+}(q,\omega
)+u_{q}^{2}A_{2\mu }(q,\omega ),\,\mu =r,a,<.  \notag
\end{eqnarray}
Here 
\begin{equation}
u_{q}^{2} = \frac{1}{2}\left( 1+\frac{\epsilon _{q}}{\sqrt{\epsilon
_{q}^{2}+\Delta _{q}^{2}}}\right), \ v_{q}^{2} = 1-u^2_q.
\end{equation}

The expressions for the retarded/advanced Green functions are given as
follows 
\begin{eqnarray}
G_{\sigma r/a}\left( k,\omega \right) &=&\left( \omega -\epsilon _{k}\pm
i0^{+}\right) ^{-1} \\
A_{lr/a}\left( q,\omega \right) &=&\left( \omega -E_{lq}\pm i0^{+}\right)
^{-1},\,l=1,2  \notag
\end{eqnarray}
and $A_{lr/a}^{+}(q,\omega )=-A_{la/r}^{\ast }(q,-\omega ).$ The
quasiparticle energies are
\begin{eqnarray}
E_{1q}=- H-\sqrt{\epsilon _{q}^{2}+\Delta_{q}^{2}}, 
\label{tunnenergy1} \\
E_{2q}=\, H-\sqrt{\epsilon _{q}^{2}+\Delta_{q}^{2}}.
\label{tunnenergy2} 
\end{eqnarray}
The distribution Green function is defined as 
\begin{equation}
G_{\sigma <}\left( k,\omega \right) =-f_{\mathrm{FD}}(\omega )\left[
G_{\sigma r}\left( k,\omega \right) -G_{\sigma a}\left( k,\omega \right) %
\right] ,
\end{equation}
$f_{\mathrm{FD}}(\omega )$ is the Fermi-Dirac function and likewise for $%
A_{l<},l=1,2.$ Next we convert the sums over the wave vectors into integrals
over the corresponding energies by introducing the density of states for the
left-hand and the right-hand side of the junction. We consider the simplest
possibility of constant density of states, that is, we do not take into
account the finite band-width effects. This choice allows us to obtain
analytical results in the zero-temperature case. The expression for the
tunneling current is obtained in the form:

I) If $\Delta _{0}\leq 2H$ 
\begin{equation}
\frac{I}{I_{0}}=
\begin{cases}
\frac{\pi }{2}\left( x_{+}^{2}-x_{-}^{2}\right) ,\hfill 0\leq eV\leq \Delta
_{0}-H \\ 
\frac{\pi }{2}\left( x_{+}^{2}-x_{-}^{2}\right) -f\left( x_{+}\right)
,\hfill \Delta _{0}-H\leq eV\leq H \\ 
\frac{\pi }{2}\left( x_{+}^{2}+x_{-}^{2}\right) -f\left( x_{+}\right)
,\hfill H\leq eV\leq \Delta _{0}+H \\ 
\frac{\pi }{2}\left( x_{+}^{2}+x_{-}^{2}\right) -f\left( x_{+}\right)
-f(x_{-}),\hfill eV\geq \Delta _{0}+H
\end{cases}
\end{equation}
II) If $\Delta _{0}\geq 2H$ 
\begin{equation}
\frac{I}{I_{0}}=
\begin{cases}
\frac{\pi }{2}\left( x_{+}^{2}-x_{-}^{2}\right) ,\hfill 0\leq eV\leq H \\ 
\frac{\pi }{2}\left( x_{+}^{2}+x_{-}^{2}\right) ,\hfill H\leq eV\leq \Delta
_{0}-H \\ 
\frac{\pi }{2}\left( x_{+}^{2}+x_{-}^{2}\right) -f\left( x_{+}\right)
,\hfill \Delta _{0}-H\leq eV\leq \Delta _{0}+H \\ 
\frac{\pi }{2}\left( x_{+}^{2}+x_{-}^{2}\right) -f\left( x_{+}\right)
-f(x_{-}),\hfill eV\geq \Delta _{0}+H
\end{cases}
\end{equation}
where $I_{0}=\pi et^{2}\rho _{N}^{(L)}\rho _{N}^{(R)}\Delta _{0},x_{\pm
}=\left( eV\pm H\right) /\Delta _{0},f(x)=x^{2}\tan ^{-1}\sqrt{x^{2}-1}-%
\sqrt{x^{2}-1}$ and $\rho _{N}^{(L,R)}$ is the density of states in the
left-hand/right-hand side of the junction. Also, we calculated the dynamical
conductance $g=dI/dV.$ The expression for it is given by

I) If $\Delta _{0}\leq 2H$ 
\begin{equation}
\frac{g}{g_{0}}=
\begin{cases}
\frac{\pi H}{\Delta _{0}},\hfill 0\leq eV\leq \Delta _{0}-H \\ 
\frac{\pi H}{\Delta _{0}}-g\left( x_{+}\right) ,\hfill \Delta _{0}-H\leq
eV\leq H \\ 
\frac{\pi eV}{\Delta _{0}}-g\left( x_{+}\right) ,\hfill H\leq eV\leq \Delta
_{0}+H \\ 
\frac{\pi eV}{\Delta _{0}}-g\left( x_{+}\right) -g(x_{-}),\hfill eV\geq
\Delta _{0}+H
\end{cases}
\end{equation}
II) If $\Delta _{0}\geq 2H$ 
\begin{equation}
\frac{g}{g_{0}}=
\begin{cases}
\frac{\pi H}{\Delta _{0}},\hfill 0\leq eV\leq H \\ 
\frac{\pi eV}{\Delta _{0}},\hfill H\leq eV\leq \Delta _{0}-H \\ 
\frac{\pi eV}{\Delta _{0}}-g\left( x_{+}\right) ,\hfill \Delta _{0}-H\leq
eV\leq \Delta _{0}+H \\ 
\frac{\pi eV}{\Delta _{0}}-g\left( x_{+}\right) -g(x_{-}),\hfill eV\geq
\Delta _{0}+H
\end{cases}
\end{equation}
where $g_{0}=2\protect\pi
e^{2}t^{2}\protect\rho _{N}^{(L)}\protect\rho _{N}^{(R)}$ and $g\left( x\right)
 =x\tan ^{-1}\sqrt{x^{2}-1}.$

The results for both the tunneling current and the differential conductance
are shown in Fig. 1 and Fig. 2. 
\begin{figure}[h]
\vspace{0.5cm} \epsfxsize=7.0cm \hspace*{-1.5cm} \epsfbox{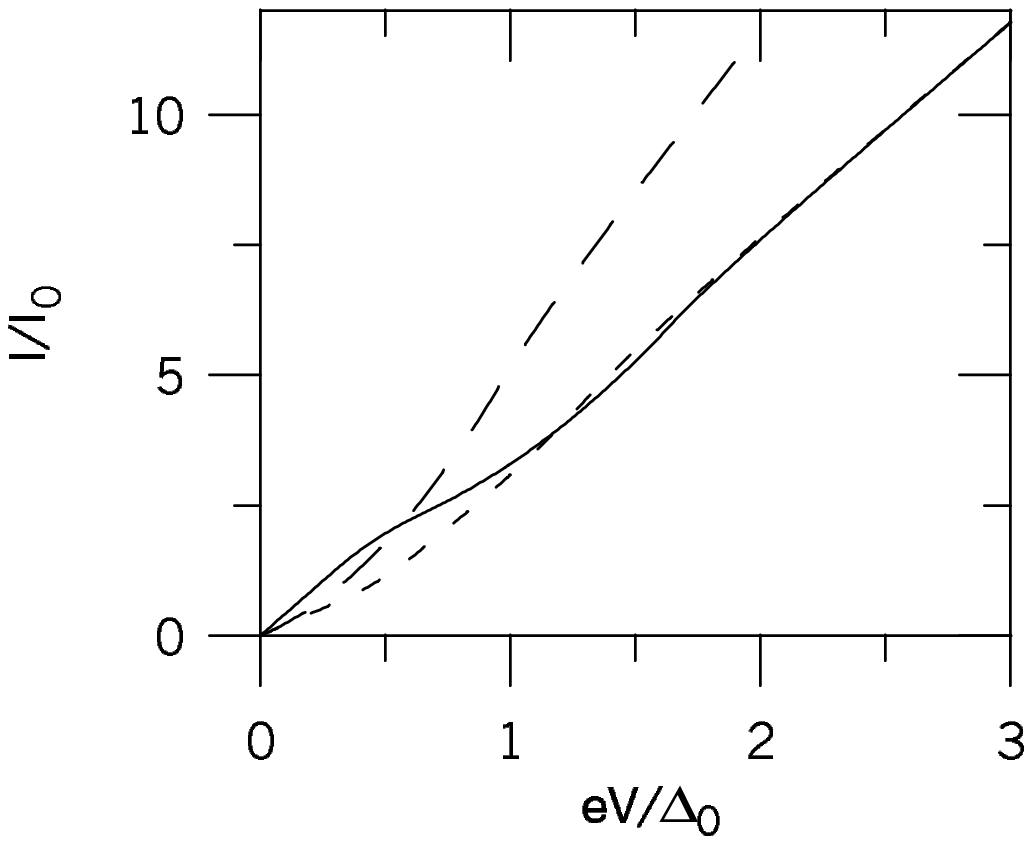}
\caption{The tunneling current as a function of the applied voltage in the
case of: i) magnon-induced superconductivity for $\Delta _{0}=1.5H$ (solid
line) and $\Delta _{0}=3H$ (short-dashed line); ii) paramagnon-induced
superconductivity (long-dashed line).}
\label{fig1}
\end{figure}
\begin{figure}[h]
\vspace{0.5cm} 
\epsfxsize=6.7cm \hspace*{-.8cm} \epsfbox{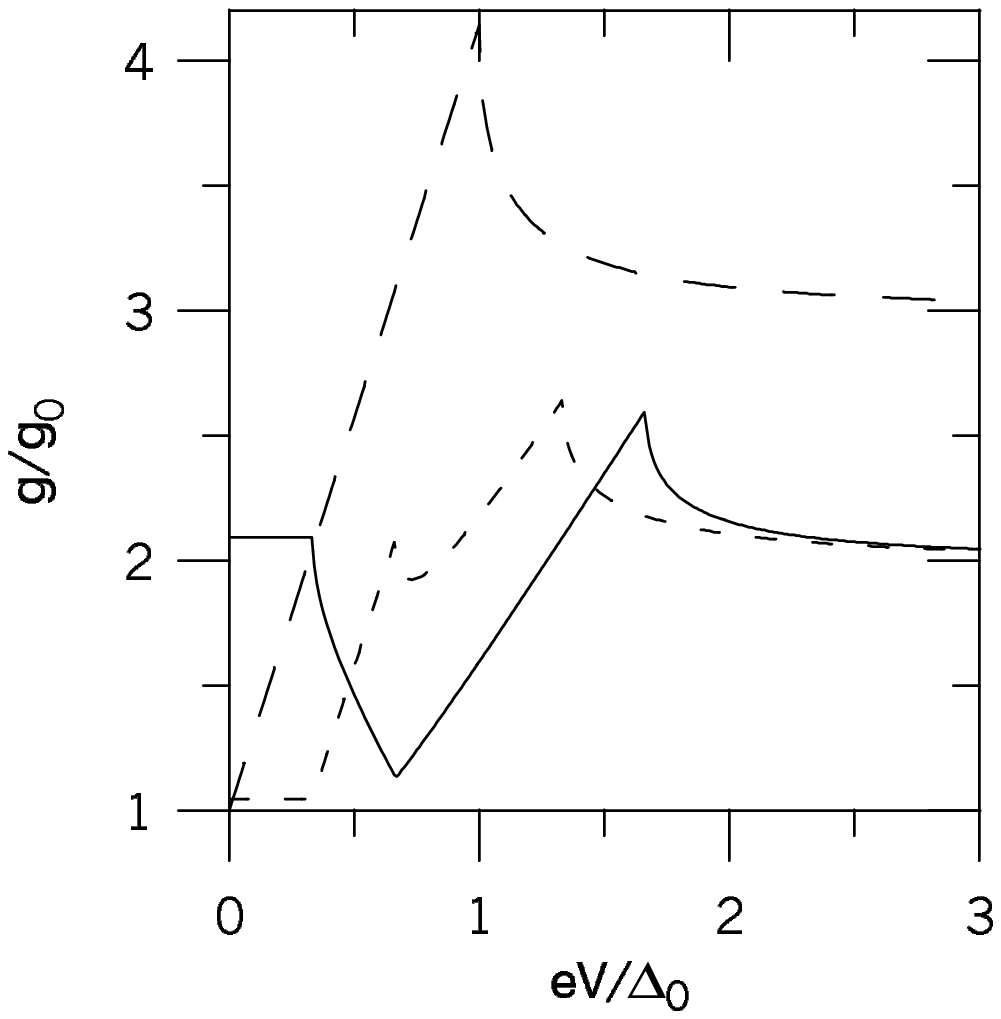} .
\caption{The differential conductance as a function of the applied voltage
in the case: i) magnon-induced superconductivity for $\Delta _{0}=1.5H$
(solid line) and $\Delta _{0}=3H$ (short-dashed line); ii)
paramagnon-induced superconductivity (long-dashed line).}
\label{fig2}
\end{figure}
The most important feature of the tunneling conductance is that it is a
continuous and smooth function of the applied voltage and does not show any
discontinuities at the gap edges that are characteristics of the tunneling
in which conventional superconductors are involved. Also, the tunneling
current is non-zero when $eV\leq \Delta _{0}$. One observes that for small
values of the applied voltage $I$ depends linearly on it, that is, the Ohm
law holds. This behavior is attributed to the existence of a Fermi surface
in this model\cite{tunn7}. Another important feature to be pointed out is
that the magnetization which is included in the Zeeman splitting energy $H$
can be measured in a tunneling experiment. When $\Delta _{0}\leq 2H$, the
differential conductance is constant for $eV\leq \Delta _{0}-H$, the minimum
is located at $eV=H$ and the maximum is at $eV=\Delta _{0}+H$ [Fig. 2, the
solid curve]. In the case of $\Delta _{0}\geq 2H,\,g$ exhibits ohmic
behavior for $eV\leq H$ and the two maximums are at $\Delta _{0}-H$ and $%
\Delta _{0}+H$ [Fig. 2, the short-dashed curve]. Thus, for any value of the
microscopic parameters one can deduce the values of both $\Delta _{0}$ and $H
$ from the tunneling conductance.

Let us consider now the paramagnon-induced superconductivity. In this case,
the tunneling current is a sum of two terms $I=I_{N}+I_{SC}$ where $I_{N}$
is the normal current due to the unpaired band of down-spin electrons and $%
I_{SC}$ is the current of the superconducting up-spin electrons. The
explicit expressions for them are written as 
\begin{eqnarray}
I_{N} &=&et^{2}\sum\limits_{k,q}\int \frac{d\omega }{2\pi }[\left[
D_{\downarrow r}\left( q,\omega \right) -D_{\downarrow a}\left( q,\omega
\right) \right] G_{\downarrow <}\left( k,\omega \right) -  \notag \\
&&D_{\downarrow <}\left( q,\omega \right) \left[ G_{\downarrow r}\left(
k,\omega \right) -G_{\downarrow a}\left( k,\omega \right) \right] ],
\end{eqnarray}
and 
\begin{eqnarray}
I_{SC} &=&et^{2}\sum\limits_{k,q}\int \frac{d\omega }{2\pi }[\left[ \Sigma_{r}
\left( q,\omega \right) -\Sigma _{a}\left( q,\omega \right) \right]
G_{\uparrow <}\left( k,\omega \right) -  \notag \\
&&\Sigma _{<}\left( q,\omega \right) \left[ G_{\uparrow r}\left( k,\omega
\right) -G_{\uparrow a}\left( k,\omega \right) \right] ],
\end{eqnarray}
where 
\begin{equation}
\Sigma _{\mu }(q,\omega )=\widetilde{u}_{q}^{2}A_{\mu }(q,\omega )+%
\widetilde{v}_{q}^{2}A_{\mu }^{+}(q,\omega ),\,\mu =r,a,<.
\end{equation}

The retarded/advanced Green function are given by $D_{\downarrow r/a}\left(
q,\omega \right) =\left( \omega -\epsilon _{q\downarrow }\pm i0^{+}\right)
^{-1},A_{r/a}\left( q,\omega \right) =\left( \omega -E_{q}\pm i0^{+}\right)
^{-1}$ and the distribution Green functions are defined as above. The
quasiparticle energy is 
\begin{equation}
E_{q}=\sqrt{\epsilon _{q\uparrow }^{2}+\Delta
_{q}^{2}}.
\label{tunnenergy3}
\end{equation} 
Also, 
\begin{equation}
\widetilde{u}_{q}^{2}=\frac{1}{2}\left( 1+\frac{\epsilon _{q\uparrow }}{E_{q}%
}\right) ,\qquad \widetilde{v}_{q}^{2}=\frac{1}{2}\left( 1-\frac{\epsilon
_{q\uparrow }}{E_{q}}\right) .
\end{equation}
With the same assumptions as above the tunneling currents are obtained in
the form 
\begin{equation}
I_{N}=2I_{0}eV/\Delta _{0}
\label{tunnnc}
\end{equation} 
\begin{equation}
I_{SC}=2I_{0}\left[ \frac{\pi }{2}\left( \frac{eV}{\Delta _{0}}\right)
^{2}-f\left( \frac{eV}{\Delta _{0}}\right) \theta \left( eV-\Delta
_{0}\right) \right] .
\end{equation}
Unlike the case of magnon-induced superconductivity the current does not
depend on the Zeeman splitting energy. The corresponding expression for the
differential conductance is cast into the form

\begin{equation}
g=g_{0}\left[ 1+\pi \frac{eV}{\Delta _{0}}-2g\left( \frac{eV}{\Delta _{0}}%
\right) \theta \left( eV-\Delta _{0}\right) \right] .
\end{equation}
The results for the tunneling current and the differential conductance are
shown in Fig. 1 and Fig. 2 (the long-dashed-line), respectively. One sees
that $g$ has no ohmic behavior and has only one maximum located at $\Delta
_{0}.$ Thus, a tunneling experiment can very easily determine the correct
pairing mechanism in the ferromagnetic superconductors.

\begin{figure}[h]
\vspace{0.5cm} \epsfxsize=7.0cm \hspace*{-1.5cm} \epsfbox{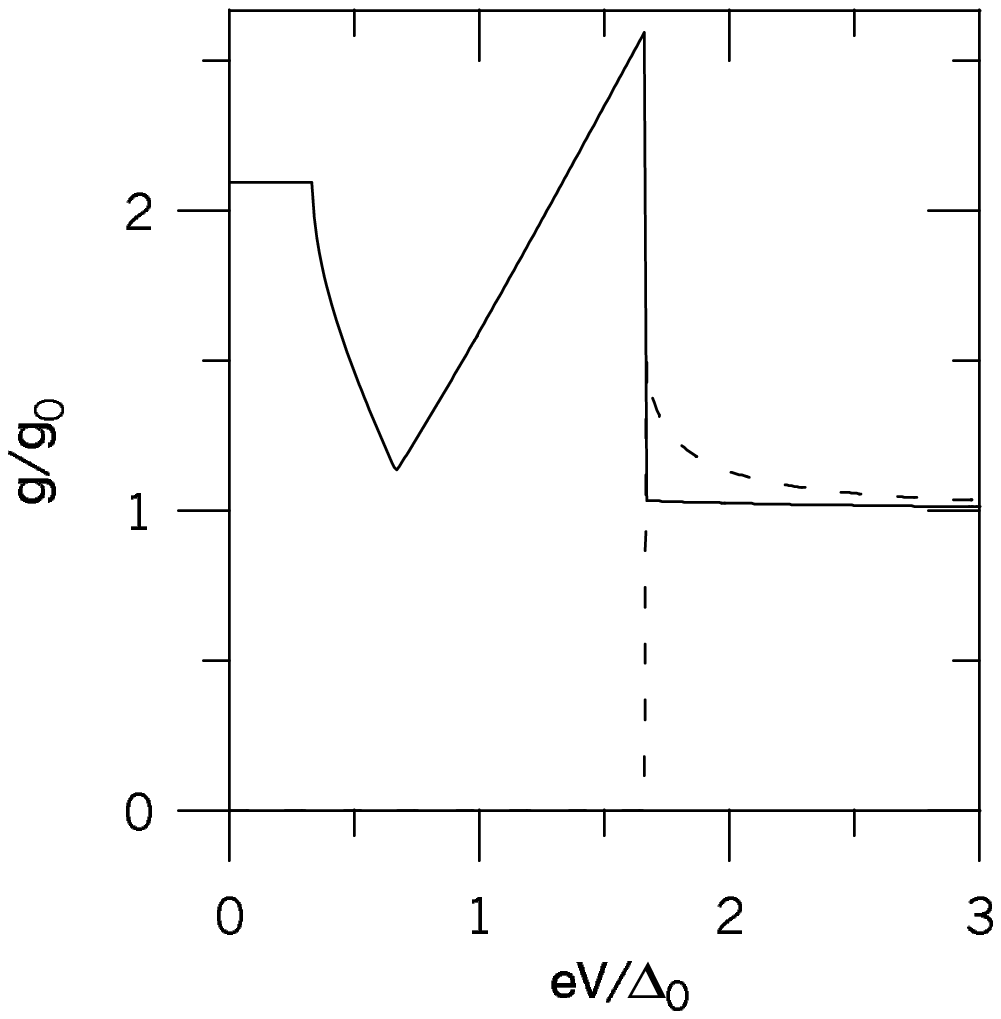}
\caption{Bogoliubov quasiparticle contribution to the differential conductance in the case of magnon-induced superconductivity for $\Delta _{0}=1.5H$: quasiparticle with energy $E_{1q}$ [Eq. (\ref{tunnenergy1})] (short-dashed
line) and quasiparticle with energy $E_{2q}$ [Eq. (\ref{tunnenergy2})] (solid line).}
\label{fig3}
\end{figure}
\begin{figure}[h]
\vspace{0.5cm} 
\epsfxsize=6.7cm \hspace*{-.8cm} \epsfbox{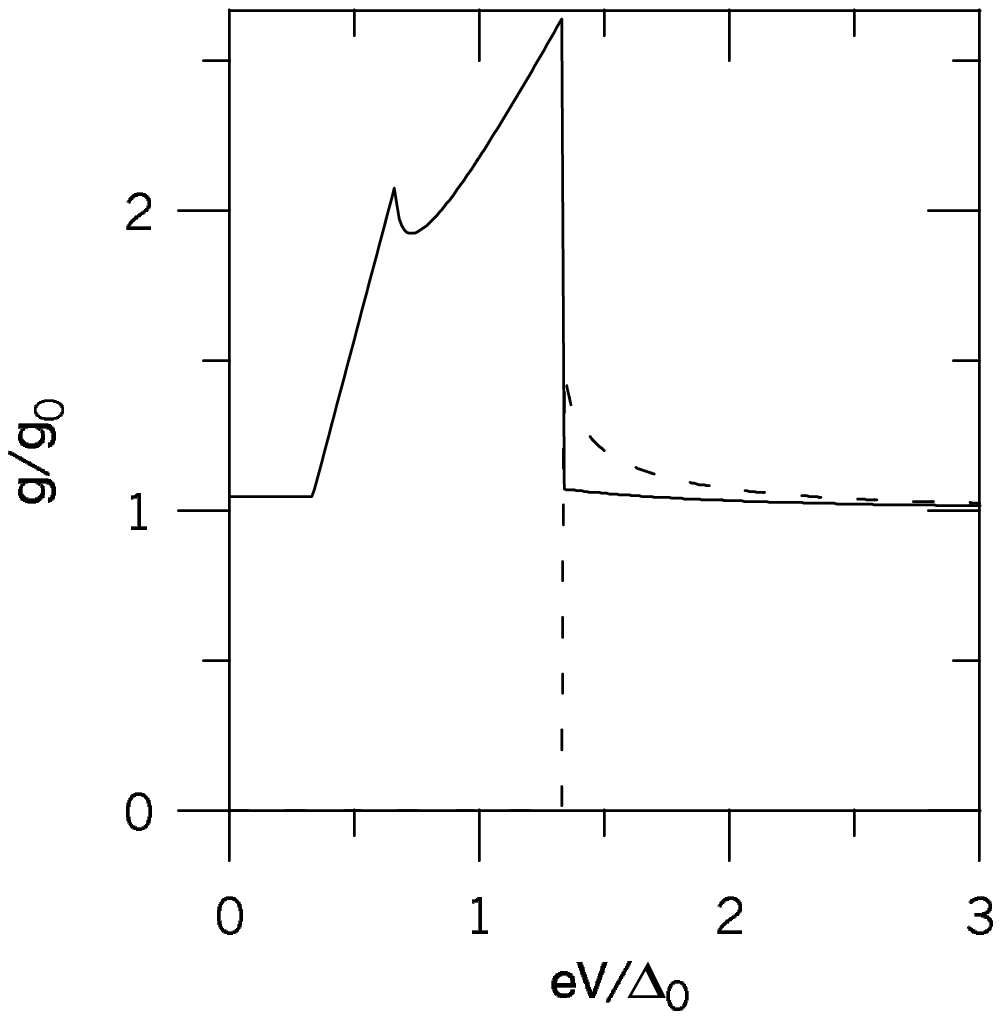} .
\caption{Bogoliubov quasiparticle contribution to the differential conductance in the case of magnon-induced superconductivity for $\Delta _{0}=3H$: quasiparticle with energy $E_{1q}$ [Eq. (\ref{tunnenergy1})] (short-dashed
line) and quasiparticle with energy $E_{2q}$ [Eq. (\ref{tunnenergy2})] (solid line).}
\label{fig4}
\end{figure}

Point-contact tunneling between normal metal and superconductor is one of
the best probes for analyzing the energy gap of superconductors.
Measurements of tunneling current yield valuable information on the symmetry
of the order parameter which in turn is essential for understanding the
mechanism of superconductivity. Our results differ from what we expect for
the conventional superconductors in many aspects. Perhaps the most striking
difference is the nonzero conductance inside the superconducting gap. The
tunneling conductance is a continuous and smooth function of the applied
voltage. For small values of the applied voltage $I$ depends linearly on it,
that is, the Ohm law holds(Fig1). The main focus in the present paper is the
evidence that the origin of this behavior is the existence of a Fermi
surface in the superconducting phase. It was experimentally observed \cite
{tunn3,tunn4} that the specific heat depends on the temperature linearly at
low temperature. This means that the superconducting state is strongly
gapless with a Fermi surface. Our theoretical prediction suggests that the
tunneling experiment could give another independent verification of the
surviving of the Fermi surface in the superconducting state.

We consider two pairing mechanisms - magnon - and paramagnon- induced
superconductivity\cite{tunn7,tunn6}. In the case of magnon induced
superconductivity the differential conductance exhibits ohmic behaviour at
low voltages for any value of the microscopic parameters, and has two
extreme points which determine the gap and the Zeeman splitting. Unlike
this, the tunneling current, in the case of paramagnon-induced
superconductivity, does not depend on the Zeeman splitting energy, the
differential conductance has no ohmic behavior, and the only local maximum
determines the gap. 

The contribution of the Bogoliubov quasiparticles to the differential conductance is shown in Fig. 3 (for $\Delta_0=1.5H$) and in Fig. 4 (for $\Delta_0=3H$) in the case of magnon-induced superconductivity. From Eqs. (\ref{tunnenergy1},\ref{tunnenergy2}) it can easily be established that there is a gap $\Delta_0+H$ in the quasiparticle energy $E_{1q}$ while the quasiparticle with energy $E_{2q}$ is a gapless excitation with a Fermi surface\cite{tunn7}. 
As a result, the contribution of the first quasiparticle to the differential conductance shows 
typical behaviour well known for gapped superconductors [Figs. 3 and 4, the short-dashed line] 
and all important features in $g$ we have already pointed out are due to the second quasiparticle
[Figs 3 and 4, the solid lines].
Namely, the ohmic part in the differential conductance at low voltage bias reflects the gapless nature
with a Fermi surface of this quasiparticle. 
The ferromagetic superconductors are anisotropic. The different symmetries of the order parameter
correspond to different anisotropies, respectively to different Fermi surfaces.As a consequence, 
calculating the tunneling current and the differential conductance and averaging over the Fermi surface 
we obtain different expressions for the different Cooper pairing mechanisms. 

It is important to compare our    results with the results in the case of isotropic s-wave superconductors. At zero magnetic field the energies of the Bogoliubov quasiparticles are degenerate and there is only one peak in the quasiparticle density of states. Correspondingly, there is only one peak in the differential conductance. The effect of the magnetic field is to lift this degeneracy. The quasiparticle peak in the density of states splits in two peaks and as a result two peaks appear in the differential conductance with each peak coming from a quasiparticle with a given spin (up or down)\cite{tunn13,tunn13b}.

In the case of magnon-induced superconductivity the quasiparticle spectrum is not degenerate too. The two extrema in the differential conductance result from a complicated averaging over the Fermi surface and reflect the symmetry of the superconducting order parameter, respectively the mechanism of the superconductivity.

In the case of paramagnon-induced superconductivity there is one gapped excitation (\ref{tunnenergy3}) which gives the only peak in the differential conductance. The other excitation is a free spin-down electron which is responsible for the linear in the bias voltage part of the tunneling current (\ref{tunnnc}).

It is known that in the case of anisotropic pairing
zero-energy bound states can exist \cite{tunn13a}. They will modify the
low-voltage behaviour of the differential conductance. The inclusion of
these states would require modification of our approach. The results we have
obtained will be changed at low voltages. However, the non-zero differential
conductance at zero voltage is a direct consequence of the existence of a
Fermi surface in the ferromagnetic superconductors. This behaviour will not
be changed qualitatively when the zero-energy bound states are taken into
account. The rest of out findings should apply in this case because they
refer to features in the differential conductance at voltages of the order
of the superconducting order parameter. The evident difference between the
tunneling processes involving magnon-induced superconductivity and those
involving paramagnon-induced superconductivity suggests that tunneling
experiment could be crucial for understanding the mechanism of ferromagnetic
superconductivity.

The ferromagnetic superconductivity was seen only in samples with small
normal-state residual electrical resistivity $\rho_0=2\mu\Omega\text{cm}$%
\cite{tunn2,tunn3}. This means that the impurity concentration is very low.
Therefore, if the impurities are taken into account they will only slightly
modify the low-voltage behaviour of the tunneling differential conductance.

 The
tunneling differential conductance in the case of s-wave superconductors in an external magnetic field has been measured long ago
\cite{tunn13}. It shows different behavior for different values of the
magnetic field [Fig. 1 in Ref.\cite{tunn13}], that is, the tunneling current measurements
are sensitive to the quasiparticle spectrum of the system in the magnetic field. This fact demonstrates that
the experiments can in principle distinguish the different behavior we
predict for the two pairing mechanisms.

Our results differ, as well, from the calculations of the tunneling
conductance in the case when spin-triplet superconductivity of different
kind is considered\cite{tunn14}. This demonstrates that by means of
tunneling experiments the symmetry of different p-wave superconductors can
be established. This is an additional support for our efforts to distinguish
the magnon and paramagnon mechanisms of FM superconductivity.



\begin{thebibliography}{*}
\bibitem[{*}]{byline}  Electronic address: tzanko@phys.uni-sofia.bg

\bibitem{tunn}  I. Giaever, Phys.Rev.Lett. \textbf{5}, 147, 464 (1960).

\bibitem{tunn1}  S. Saxena, P. Agarwal, K. Ahilan, F. M. Grosche, R.
Haselwimmer, M. Steiner, E. Pugh, I. Walker, S. Julian, P. Monthoux, G.
Lonzarich, A. Huxley, I. Sheikin, D. Braithwaite, and J. Flouquet, Nature
(London) \textbf{406}, 587 (2000).

\bibitem{tunn2}  D. Aoki, A. Huxley, E. Ressouche, D. Braithwaite, J.
Flouquet, J-P. Brison, E.Lhotel, and C. Paulsen, Nature (London) \textbf{413}%
, 613 (2001).

\bibitem{tunn3}  C. Pfleiderer, M. Uhlarz, S. Hayden, R. Vollmer, H.v. L\"{o}%
hneysen, N. Bernhoeft, and G. Lonzarich, Nature (London) \textbf{412}, 58
(2001).

\bibitem{tunn4}  A. Huxley, I. Sheikin, E. Ressouche, N. Kernavanois, D.
Braithwaite, R. Calemczuk, and J. Flouquet, Phys. Rev. B \textbf{63}, 144519
(2001).

\bibitem{tunn5}  N. Tateiwa, T. Kobayashi, K. Hanazono, K. Amaya, Y. Haga,
R. Settai, and Y. Onuki, J. Phys. Condens. Matter \textbf{13}, L17 (2001).
Phys. Rev. B \textbf{63}, 144519 (2001).

\bibitem{tunn6}  D. Fay and J. Apple, Pys.Rev. B \textbf{22}, 3173 (1980).

\bibitem{tunn7}  N. Karchev, Phys. Rev. \textbf{B 67}, 054416 (2003); N.
Karchev, J. Phys.Condens.Matter, \textbf{15}, L385 (2003).

\bibitem{tunn8}  K. Machida and T. Ohmi, Phys. Rev. Lett. \textbf{86}, 850
(2001).

\bibitem{tunn9}  J. Bardeen, Phys.Rev.Lett. \textbf{6}, 57 (1961).

\bibitem{tunn10}  M. H. Cohen, L. M. Falicov, and J. C. Phillips, Phys. Rev.
Lett. \textbf{8}, 316 (1962).

\bibitem{tunn10a}  K. Ueda and T. M. Rice, Phys. Rev. \textbf{B31}, 7114
(1985).

\bibitem{tunn11}  Kuang-chao Chou, Zhao-bin Su, Bai-lin Hao, and Lu Yu,
Phys. Rep. \textbf{118}, 1 (1985).

\bibitem{tunn12}  G. D. Mahan, \textit{Many-Particle Physics} (Plenum, New
York, 1981), Ch. 9.

\bibitem{tunn13}  R. Meservey, P. M. Tedrow, and P. Fulde, Phys. Rev. Lett. 
\textbf{25}, 1270 (1970).

\bibitem{tunn13b} R. Meservey, P. M. Tedrow, Phys. Reports, \textbf{238}, 175 (1994)

\bibitem{tunn13a}  R. Joynt, J. Low Temp. Physics \textbf{109}, 811 (1997).

\bibitem{tunn14}  C.J.Bolech and T.Giamarchi, Phys. Rev. Lett. \textbf{92},
127001 (2004)
\end{thebibliography}
\end{document}